\documentclass[12pt]{article}

\usepackage[dvipdfmx]{graphicx}
\usepackage[dvipdfmx]{xcolor}
\usepackage{amsmath}
\usepackage{amssymb}
\usepackage{tikz}
\usepackage{cite}
\usepackage{fancyhdr}

\def\erase#1{{}}
\def\EqArrerase#1{{}}

\makeatletter
 \renewcommand{\theequation}{%
 \thesection.\arabic{equation}}
 \@addtoreset{equation}{section}
\makeatother

\def\T{{\rm T}}

\def\GL{{G\kern-.12em L\kern-.04em}}
\def\OSp{{O\kern-.11em S\kern-.04em p}}
\def\IOSp{{I\kern-.06em O\kern-.11em S\kern-.04em p}}
\def\MN{{M\kern-.14em N}}
\def\NM{{N\kern-.14em M}}
\def\NL{{N\kern-.14em L}}
\def\LN{{L\kern-.11em N}}
\def\ML{{M\kern-.14em L}}
\def\LM{{L\kern-.11em M}}
\def\RN{{R\kern-.11em N}}
\def\NR{{N\kern-.14em R}}
\def\RM{{R\kern-.11em M}}
\def\MR{{M\kern-.14em R}}
\def\RL{{R\kern-.11em L}}
\def\LR{{L\kern-.11em R}}
\def\RS{{R\kern-.11em S}}
\def\SR{{S\kern-.11em R}}
\def\SN{{S\kern-.11em N}}
\def\NS{{N\kern-.11em S}}
\def\SM{{S\kern-.11em M}}
\def\MS{{M\kern-.11em S}}
\def\SL{{S\kern-.11em L}}
\def\LS{{L\kern-.11em S}}
\def\sqr#1#2{{\vcenter{\hrule height.#2pt
      \hbox{\vrule width.#2pt height#1pt \kern#1pt
          \vrule width.#2pt}
      \hrule height.#2pt}}}
\def\bra0{\langle0|}
\def\ket0{|0\rangle}
\def\soeji#1_#2#3{#1_{#2}\cdots#1_{#3}}
\def\longgLRarrow{\longleftarrow\kern-3pt\relbar\kern-3pt\relbar\kern-3pt%
\longrightarrow}
\def\longLRarrow{\longleftarrow\kern-3pt\relbar\kern-3pt\longrightarrow}
\def\longLarrow{\longleftarrow\kern-3pt\relbar\kern-3pt\relbar\kern-3pt\relbar}
\def\longRarrow{\relbar\kern-3pt\relbar\kern-3pt\relbar\kern-3pt\longrightarrow}
\def\bothDer#1#2#3{%
\overset{\kern-.7em\stackrel{#1}{#2}}{\partial_{#3}}}
 
\makeatletter
 \renewcommand{\theequation}{%
 \thesection.\arabic{equation}}
 \@addtoreset{equation}{section}
\makeatother
\begin{document}
\thispagestyle{fancy}

\title{Bound States in Scalar Theory with Fourth-order Derivative Term}

\author{Ichiro Oda
\footnote{Electronic address: ioda@cs.u-ryukyu.ac.jp}
\\
{\it\small
\begin{tabular}{c}
Department of Physics, Faculty of Science, University of the 
           Ryukyus,\\
           Nishihara, Okinawa 903-0213, Japan\\      
\end{tabular}
}
}
\date{}

\maketitle

\thispagestyle{fancy}

\begin{abstract}

We consider a problem whether bound states are made in a scalar theory with a fourth-order derivative 
term or not. After rewriting the theory to a standard scalar theory with second-order derivative terms,
we calculate a correlation function of the composite operators made out of massive ghosts with negative norm
and massive normal fields with positive norm in the ladder approximation. It is shown that there appears a pole of the bound state 
in the correlation function of the both fields by attraction due to scalar field when the coupling constant is large whereas
there does not so in the correlation function of almost massless normal particles corresponding to the graviton. We also point out the
relationship between the scalar theory with a fourth-order derivative term and quadratic gravity. Our model 
may shed some light on the confinement of massive ghost in quadratic gravity, thereby enabling us to solve the problem 
of unitarity violation associated with the massive ghost.
     
\end{abstract}

\newpage
\pagestyle{plain}
\pagenumbering{arabic}
%\setcounter{page}{1}

%%%%%%%%%%%%%%%%%%%%%%%%%%%%%%%%%%%%%%%%%%%%%%%%%%%%%%%%%%%%%%%%%%
%%%%%%%%%%%%%%%%%%%%%%%% Article %%%%%%%%%%%%%%%%%%%%%%%%%%%%%%%%%
%%%%%%%%%%%%%%%%%%%%%%%%%%%%%%%%%%%%%%%%%%%%%%%%%%%%%%%%%%%%%%%%%%

\section{Introduction}

Quantum field theory (QFT) can be arguably regarded as the pinnacle of modern science, but QFT in its present form
is still incomplete because of several reasons. For instance, almost nothing is known of the strong coupling regime and
gravity is not quantized in a consistent way.   

One of restrictions in QFT is that the action is made out of two powers of the derivative, which goes back to the works by
Newton. In modern applications of field theory to problems far beyond particle physics, there is no restriction at all to
impose this restriction \cite{Zee}. These field theories are formulated in the Euclidean signature so the corresponding
functional integral, which have higher derivative terms, certainly makes sense. It is field theories formulated in the Minkowskian
signature that we do not know how to handle higher derivatives.

In an attempt of the construction of quantum gravity, it is known that one can establish a renormalizable quantum theory 
of gravity by adding the square terms of the scalar curvature and the Ricci tensor (or instead, the conformal tensor) with four
powers of the derivative to the Einstein-Hilbert term with two powers of the derivative in a classical action, which is nowadays 
called ${\it{quadratic \, gravity}}$, except for a critical failure of the unitarity violation arising from ${\it{massive \, ghost}}$ \cite{Stelle}. 

Incidentally, it is of interest to note that quadratic gravity is asymptotically free in a wide range of parameter regions 
\cite{Mario, Fradkin, Avramidi} while Einstein's general relativity is asymptotically non-free. By the analogy with quantum 
chromodynamics (QCD) and quantum electrodynamics (QED), it seems to be natural to conjecture that a kind of confinement 
might happen in quadratic gravity as in QCD. If so, what particle is confined? Since the graviton emerges from the Einstein-Hilbert term 
and is an essential building block in any gravitational theory, we expect that the spin $2$ massive ghost 
might be confined like quarks and gluons in QCD, enabling us to resolve the problem of the unitarity violation arising from 
the massive ghost. 

In this article, we would like to attack the problem of the unitarity violation associated with the massive ghost from the
viewpoint of the confinement of the massive ghost. However, quadratic gravity is a rather complicated theory in the sense that 
the fields carry various tensor indices and the action has an infinite number of interaction terms etc., so we wish to 
deal with the problem of the massive ghost in a much simpler scalar theory with a higher derivative term where there are 
no tensor indices and there is only a single $\Phi^4$ interaction term. 

In this simplified scalar theory, there is an interesting feature common to quadratic gravity in that in the both theories
the interaction is not repulsive but attractive. In this context, it is useful to recall that the exchange of a spin $0$ produces
an attractive force, of a spin $1$ particle a repulsive force, and of a spin $2$ particle an attractive force, for instance, as seen 
in the Yukawa pion theory, the electromagnetic theory and the Einstein gravity, respectively. Moreover, the both theories
have the massive ghost stemming from the higher derivative term and very light normal particle. Of course,
the light normal particle in the scalar theory corresponds to the graviton in quadratic gravity and must be strictly massless 
due to the general coordinate symmetry. 

On the other hand, there is a critical disadvantage in the scalar theory in the flat Minkowski background in that 
there is no general coordinate transformation (GCT) in a curved space-time. Owing to lack of the GCT, or more properly speaking, 
no BRST symmetry corresponding to the GCT at the quantum level, in this article we cannot present the whole idea about 
confinement of the massive ghost, thereby making the problem of the unitarity violation nulltify. Thus here, as the prelude 
to a resolution to the problem of the massive ghost in future, we will simply show that a massive ghost in the scalar theory 
with fourth-order derivative term makes a bound state in the ladder approximation. In this connection, recall that there appear many of bound states 
in the confinement phase of QCD.       

In a sense, our theory might be regarded as a scalar analog of the Nambu-Jona-Lasinio theory on dynamical mass generation
or spontaneous symmetry breakdown of the chiral symmetry \cite{Nambu}. However, unlike the Nambu-Jona-Lasinio theory, 
our theory is a renormalizable theory so we do not have to handle the highly divergent quantities by introducing an ${\it{ad \, hoc}}$
relativistic cutoff or form factor in actual calculations. Thus our theory may also be regarded as an approximate treatment
of the bound state model under the ladder approximation without cutoff.      

The paper is organized as follows: In the next section, we review a scalar theory with fourth-order derivative term and rewrite it into
the standard scalar theory with second-order derivative terms. In Section 3, we introduce our model and present its canonical formalism.
In Section 4, we calculate the correlation function of a composite operator. In Section 5 we analyze the pole equation. 
In Section 6, we point out the relationship between our scalar model with the fourth-order derivative term and quadratic gravity.
In final section, we draw our conclusion. We put the Appendix which accounts for a renormalization of the composite operator.

\section{Scalar theory with fourth-order derivative term}

We would like to review a scalar theory with the fourth-order derivative term \cite{Creminelli}. We follow the presentation 
by Ref. \cite{Grinstein}.

Let us start with a classical Lagrangian density of a generic scalar theory with the fourth-order derivative term: 
%**   Lag0   %%%%%%%%%%%%%%%%%%%%%%%%%%%%%%%%%%%%%%%%%%%%%%%%%%%%%%%%%
\begin{eqnarray}
{\cal{L}}_0 &=& - \frac{1}{2} ( \partial_\mu \Phi )^2 - \frac{1}{2} m^2 \Phi^2 - \frac{1}{2 M^2} ( \Box \Phi )^2 + {\cal{L}}_{\rm{int}},
\nonumber\\
&=& \frac{1}{2} \Phi \left( \Box - m^2 - \frac{1}{M^2} \Box^2 \right) \Phi + {\cal{L}}_{\rm{int}},
\label{Lag0}  
\end{eqnarray}
%%%%%%%%%%%%%%%%%%%%%%%%%%%%%%%%%%%%%%%%%%%%%%%%%%%%%%%%%%%%%%%%%%% 
where $m$ and $M$ are constants of mass dimension, and ${\cal{L}}_{\rm{int}}$ denotes an interaction Lagrangian density.\footnote{We 
take the metric convention $\eta_{\mu\nu} = \rm{diag} ( -1, +1, +1, +1)$ and define $\Box \equiv \eta_{\mu\nu} \partial^\mu \partial^\nu$.} 

From this Lagrangian density, the propagator of the scalar field $\Phi$ in momentum space reads
%**   Prop-Phi   %%%%%%%%%%%%%%%%%%%%%%%%%%%%%%%%%%%%%%%%%%%%%%%%%%%%%%%%%
\begin{eqnarray}
D_{\Phi} = \frac{1}{p^2 + \frac{p^4}{M^2} + m^2},
\label{Prop-Phi}  
\end{eqnarray}
%%%%%%%%%%%%%%%%%%%%%%%%%%%%%%%%%%%%%%%%%%%%%%%%%%%%%%%%%%%%%%%%%%% 
from which the poles are determined by the equation
%**   Pole-Phi   %%%%%%%%%%%%%%%%%%%%%%%%%%%%%%%%%%%%%%%%%%%%%%%%%%%%%%%%%
\begin{eqnarray}
p^2 + \frac{p^4}{M^2} + m^2 = 0.
\label{Pole-Phi}
\end{eqnarray}
%%%%%%%%%%%%%%%%%%%%%%%%%%%%%%%%%%%%%%%%%%%%%%%%%%%%%%%%%%%%%%%%%%% 
In order that the propagator (\ref{Prop-Phi}) describes two physical degrees of freedom with positive mass, the pole equation (\ref{Pole-Phi})
must have two real solutions so we have a reality condition on the solutions:
%**   Real-cond   %%%%%%%%%%%%%%%%%%%%%%%%%%%%%%%%%%%%%%%%%%%%%%%%%%%%%%%%%
\begin{eqnarray}
D \equiv M^4 - 4 M^2 m^2 = M^2 ( M^2 - 4 m^2 ) > 0.
\label{Real-cond}  
\end{eqnarray}
%%%%%%%%%%%%%%%%%%%%%%%%%%%%%%%%%%%%%%%%%%%%%%%%%%%%%%%%%%%%%%%%%%% 
We therefore have an inequality:
%**   Real-cond2   %%%%%%%%%%%%%%%%%%%%%%%%%%%%%%%%%%%%%%%%%%%%%%%%%%%%%%%%%
\begin{eqnarray}
M >  2 m.
\label{Real-cond2}  
\end{eqnarray}
%%%%%%%%%%%%%%%%%%%%%%%%%%%%%%%%%%%%%%%%%%%%%%%%%%%%%%%%%%%%%%%%%%% 
This condition will also appear shortly in taking account of the diagonalization of two scalar fields. For $M \gg m$, we have two poles
at $p^2 \simeq - m^2$ and $p^2 \simeq - M^2$. 
 
To rewrite the fourth-derivative term to the usual second-derivative term, it is useful to introduce an auxiliary scalar field $\eta$:
%**   Lag1   %%%%%%%%%%%%%%%%%%%%%%%%%%%%%%%%%%%%%%%%%%%%%%%%%%%%%%%%%
\begin{eqnarray}
{\cal{L}} &=& - \frac{1}{2} ( \partial_\mu \Phi )^2 - \frac{1}{2} m^2 \Phi^2 + \eta \Box \Phi + \frac{1}{2} M^2 \eta^2 + {\cal{L}}_{\rm{int}}.
\label{Lag1}  
\end{eqnarray}
%%%%%%%%%%%%%%%%%%%%%%%%%%%%%%%%%%%%%%%%%%%%%%%%%%%%%%%%%%%%%%%%%%% 
Actually, the path integration over $\eta$ produces the Lagrangian density (\ref{Lag0}) so the two Lagrangian densities are quantum mechanically
equivalent.  Next, instead of the scalar field $\Phi$, let us introduce a new scalar field $\omega$ which is defined in terms of a linear
combination 
%**   Omega   %%%%%%%%%%%%%%%%%%%%%%%%%%%%%%%%%%%%%%%%%%%%%%%%%%%%%%%%%
\begin{eqnarray}
\omega = \Phi + \eta,
\label{Omega}  
\end{eqnarray}
%%%%%%%%%%%%%%%%%%%%%%%%%%%%%%%%%%%%%%%%%%%%%%%%%%%%%%%%%%%%%%%%%%% 
by which the Lagrangian density (\ref{Lag1}) can be cast to the form: 
%**   Lag2   %%%%%%%%%%%%%%%%%%%%%%%%%%%%%%%%%%%%%%%%%%%%%%%%%%%%%%%%%
\begin{eqnarray}
{\cal{L}} &=& - \frac{1}{2} ( \partial_\mu \omega )^2 - \frac{1}{2} m^2 \omega^2 + \frac{1}{2} ( \partial_\mu \eta )^2 
+ \frac{1}{2} ( M^2 - m^2 ) \eta^2 
\nonumber\\
&+& m^2 \omega \eta + {\cal{L}}_{\rm{int}}.
\label{Lag2}  
\end{eqnarray}
%%%%%%%%%%%%%%%%%%%%%%%%%%%%%%%%%%%%%%%%%%%%%%%%%%%%%%%%%%%%%%%%%%% 

There is a mixing term, $m^2 \omega \eta$, but one can diagonalize it by performing a sympletic rotation:
%**   Sympletic   %%%%%%%%%%%%%%%%%%%%%%%%%%%%%%%%%%%%%%%%%%%%%%%%%%%%%%%%%
\begin{eqnarray}
\left(
\begin{array}{r}
\omega \\
\eta \\
\end{array}
\right) =
\left(
\begin{array}{rr}
\cosh \theta & \sinh \theta \\
\sinh \theta & \cosh \theta \\
\end{array}
\right)
\left(
\begin{array}{r}
\phi \\
\varphi \\
\end{array}
\right).
\label{Sympletic}  
\end{eqnarray}
%%%%%%%%%%%%%%%%%%%%%%%%%%%%%%%%%%%%%%%%%%%%%%%%%%%%%%%%%%%%%%%%%%% 
The result is given by
%**   Lag3   %%%%%%%%%%%%%%%%%%%%%%%%%%%%%%%%%%%%%%%%%%%%%%%%%%%%%%%%%
\begin{eqnarray}
{\cal{L}} = - \frac{1}{2} ( \partial_\mu \phi )^2 - \frac{1}{2} m_\phi^2 \phi^2 + \frac{1}{2} ( \partial_\mu \varphi )^2 
+ \frac{1}{2} m_\varphi^2 \varphi^2 + {\cal{L}}_{\rm{int}}.
\label{Lag3}  
\end{eqnarray}
%%%%%%%%%%%%%%%%%%%%%%%%%%%%%%%%%%%%%%%%%%%%%%%%%%%%%%%%%%%%%%%%%%% 
Here the rotation angle $\theta$ is taken to obey the relation:
%**   Theta   %%%%%%%%%%%%%%%%%%%%%%%%%%%%%%%%%%%%%%%%%%%%%%%%%%%%%%%%%
\begin{eqnarray}
\tanh 2 \theta = \frac{2 m^2}{2 m^2 - M^2}.
\label{Theta}  
\end{eqnarray}
%%%%%%%%%%%%%%%%%%%%%%%%%%%%%%%%%%%%%%%%%%%%%%%%%%%%%%%%%%%%%%%%%%% 
Note that this relation naturally leads to the inequality (\ref{Real-cond2}). As for the squared masses, $m_\phi^2$ and $m_\varphi^2$ are
defined as
%**   2-mass   %%%%%%%%%%%%%%%%%%%%%%%%%%%%%%%%%%%%%%%%%%%%%%%%%%%%%%%%%
\begin{eqnarray}
m_\phi^2 &=& m^2 ( \cosh \theta - \sinh \theta )^2 - M^2 \sinh^2 \theta,
\nonumber\\
m_\varphi^2 &=& - m^2 ( \cosh \theta - \sinh \theta )^2 + M^2 \cosh^2 \theta,
\label{2-mass}  
\end{eqnarray}
%%%%%%%%%%%%%%%%%%%%%%%%%%%%%%%%%%%%%%%%%%%%%%%%%%%%%%%%%%%%%%%%%%% 
whose magnitudes depend on $\theta$ and can take any positive values as long as they satisfy the relation
%**   mass-rel   %%%%%%%%%%%%%%%%%%%%%%%%%%%%%%%%%%%%%%%%%%%%%%%%%%%%%%%%%
\begin{eqnarray}
m_\phi^2 + m_\varphi^2 = M^2.
\label{mass-rel}  
\end{eqnarray}
%%%%%%%%%%%%%%%%%%%%%%%%%%%%%%%%%%%%%%%%%%%%%%%%%%%%%%%%%%%%%%%%%%% 
Keeping the application to quadratic gravity in mind, $m_\phi$ should be very small (ideally speaking, vanishing) and $m_\varphi$ be of the same 
order as the Planck mass, so in this article we will assume that $\theta \approx 0$ and $M \gg m$. Then we have 
$m_\phi \approx m, \, m_\varphi \approx M$ and $m_\phi \ll m_\varphi$.   
  
Finally, let us specify the form of the interaction term. For the sake of simplicity, we shall take the quartic interaction:
%**   Quartic-int   %%%%%%%%%%%%%%%%%%%%%%%%%%%%%%%%%%%%%%%%%%%%%%%%%%%%%%%%%
\begin{eqnarray}
{\cal{L}}_{\rm{int}} = - \frac{f}{4 !} \Phi^4,
\label{Quartic-int}  
\end{eqnarray}
%%%%%%%%%%%%%%%%%%%%%%%%%%%%%%%%%%%%%%%%%%%%%%%%%%%%%%%%%%%%%%%%%%% 
which can be written in terms of $\phi$ and $\varphi$ fields as 
%**   Quartic-int2   %%%%%%%%%%%%%%%%%%%%%%%%%%%%%%%%%%%%%%%%%%%%%%%%%%%%%%%%%
\begin{eqnarray}
{\cal{L}}_{\rm{int}} = - \frac{\lambda}{4 !} ( \phi - \varphi )^4,
\label{Quartic-int2}  
\end{eqnarray}
%%%%%%%%%%%%%%%%%%%%%%%%%%%%%%%%%%%%%%%%%%%%%%%%%%%%%%%%%%%%%%%%%%% 
where the coupling constant $\lambda$ is defined by   
%**   Coupling-lambda   %%%%%%%%%%%%%%%%%%%%%%%%%%%%%%%%%%%%%%%%%%%%%%%%%%%%%%%%%
\begin{eqnarray}
\lambda \equiv f ( \cosh \theta - \sinh \theta )^4.
\label{Coupling-lambda}  
\end{eqnarray}
%%%%%%%%%%%%%%%%%%%%%%%%%%%%%%%%%%%%%%%%%%%%%%%%%%%%%%%%%%%%%%%%%%% 
In case of $M \gg m$, this coupling constant is approximated by $f$, i.e., $\lambda \approx f$.

\section{Our model}

In this article, compared to the Lagrangian density (\ref{Lag3}) plus (\ref{Quartic-int2}), we will use a slightly generalized model 
whose Lagrangian density is of form:
%**   Start-Lag   %%%%%%%%%%%%%%%%%%%%%%%%%%%%%%%%%%%%%%%%%%%%%%%%%%%%%%%%%
\begin{eqnarray}
{\cal{L}} = - \frac{1}{2} ( \partial_\mu \phi )^2 - \frac{1}{2} m_\phi^2 \phi^2 - \epsilon \left[  \frac{1}{2} ( \partial_\mu \varphi )^2 
+ \frac{1}{2} m_\varphi^2 \varphi^2 \right] - \frac{\lambda}{4 !} ( \phi - \varphi )^4,
\label{Start-Lag}  
\end{eqnarray}
%%%%%%%%%%%%%%%%%%%%%%%%%%%%%%%%%%%%%%%%%%%%%%%%%%%%%%%%%%%%%%%%%%% 
where we have put the sign factor $\epsilon$ in front of the third term on the right hand side (RHS). Because of it, $\varphi$ is a ghost 
with negative norm or a normal particle with positive norm depending on $\epsilon = -1$ or $\epsilon = +1$, respectively. In Ref. \cite{KK2}, 
based on the similar Lagrangian density to this one, vertex functions have been investigated by using $1/N$-expansion of normal ``matter fields'' 
$\phi_i \, (i = 1, 2, \dots, N)$ but since the Lagrangian density at hand has only one normal field $\phi$ except for the case of $\epsilon = +1$, 
we cannot rely on the $1/N$-expansion. Since we have in mind in future that our theory should apply to quadratic gravity where the two fields 
$\phi$ and $\varphi$, respectively correspond to the graviton and the massive ghost, we will not make use of the $1/N$-expansion.     

It is now straightforward to make the canonical quantization by following the standard recipe:
%**   Can-quant   %%%%%%%%%%%%%%%%%%%%%%%%%%%%%%%%%%%%%%%%%%%%%%%%%%%%%%%%%
\begin{eqnarray}
\phi (x) &=& \int \frac{d^3 q}{\sqrt{(2 \pi)^3 2 q^0}} \left[ a_\phi (\Vec{q}) e^{i q x} +  a_\phi^\dagger (\Vec{q}) e^{- i q x} \right],
\nonumber\\
\varphi (x) &=& \int \frac{d^3 q}{\sqrt{(2 \pi)^3 2 q^0}} \left[ a_\varphi (\Vec{q}) e^{i q x} + a_\varphi^\dagger (\Vec{q}) e^{- i q x} \right],
\label{Can-quant}  
\end{eqnarray}
%%%%%%%%%%%%%%%%%%%%%%%%%%%%%%%%%%%%%%%%%%%%%%%%%%%%%%%%%%%%%%%%%%% 
where $q x = - q^0 x^0 + \Vec{q} \Vec{x} = - \sqrt{\Vec{q}\,^2 + m_a^2} + \Vec{q} \Vec{x}$ and $m_a = (m_\phi, m_\varphi)$. 
The commutation relations read:
%**   CRs   %%%%%%%%%%%%%%%%%%%%%%%%%%%%%%%%%%%%%%%%%%%%%%%%%%%%%%%%%
\begin{eqnarray}
&{}& [ a_\phi (\Vec{p}), a_\phi^\dagger (\Vec{q}) ] = \delta^3 (p-q),   
\nonumber\\
&{}& [ a_\varphi (\Vec{p}), a_\varphi^\dagger (\Vec{q}) ] = \epsilon \delta^3 (p-q). 
\label{CRs}  
\end{eqnarray}
%%%%%%%%%%%%%%%%%%%%%%%%%%%%%%%%%%%%%%%%%%%%%%%%%%%%%%%%%%%%%%%%%%% 

From these commutation relations, it is easy to derive the propagators at the tree level:
%**   Prop-phi   %%%%%%%%%%%%%%%%%%%%%%%%%%%%%%%%%%%%%%%%%%%%%%%%%%%%%%%%%
\begin{eqnarray}
D_\phi (x-y) &\equiv& \langle 0 | T \phi(x) \phi(y) | 0 \rangle
\nonumber\\
&=& \int \frac{d^4 q}{i (2 \pi)^4} e^{i q \cdot (x - y)} \frac{1}{q^2 + m_\phi^2 - i \varepsilon}
\nonumber\\
&\equiv& \int \frac{d^4 q}{i (2 \pi)^4} e^{i q \cdot (x - y)} D_\phi (q),
\label{Prop-phi}  
\end{eqnarray}
%%%%%%%%%%%%%%%%%%%%%%%%%%%%%%%%%%%%%%%%%%%%%%%%%%%%%%%%%%%%%%%%%%% 
and 
%**   Prop-varphi   %%%%%%%%%%%%%%%%%%%%%%%%%%%%%%%%%%%%%%%%%%%%%%%%%%%%%%%%%
\begin{eqnarray}
D_\varphi (x-y) &\equiv& \langle 0 | T \varphi(x) \varphi(y) | 0 \rangle
\nonumber\\
&=& \int \frac{d^4 q}{i (2 \pi)^4} e^{i q \cdot (x - y)} \frac{\epsilon}{q^2 + m_\varphi^2 - i \varepsilon}
\nonumber\\
&\equiv& \int \frac{d^4 q}{i (2 \pi)^4} e^{i q \cdot (x - y)} D_\varphi (q).
\label{Prop-varphi}  
\end{eqnarray}
%%%%%%%%%%%%%%%%%%%%%%%%%%%%%%%%%%%%%%%%%%%%%%%%%%%%%%%%%%%%%%%%%%% 
Here we have defined
%**   D-mom   %%%%%%%%%%%%%%%%%%%%%%%%%%%%%%%%%%%%%%%%%%%%%%%%%%%%%%%%%
\begin{eqnarray}
D_\phi (q) = \frac{1}{q^2 + m_\phi^2 - i \varepsilon},    \qquad
D_\varphi (q) = \frac{\epsilon}{q^2 + m_\varphi^2 - i \varepsilon}. 
\label{D-mom}  
\end{eqnarray}
%%%%%%%%%%%%%%%%%%%%%%%%%%%%%%%%%%%%%%%%%%%%%%%%%%%%%%%%%%%%%%%%%%% 
In the scalar field theory with $\Phi^4$ interaction used in this article, the two point function at the next leading level arises from a two-loop
graph, so it will not be convenient to use this theory to illustrate the calculation of the two point function.

\section{Correlation function of composite operator}

In this section, we will derive the correlation function of a composite operator of scalar fields
on the basis of the canonical operator formalism explained in the previous section. 
The similar problem of the Lee's complex ghost model \cite{LW1, LW2, Lee, Nakanishi1, KK1} has been recently discussed in Refs. \cite{Asorey1, Asorey2} 
in the path integral approach and in Ref. \cite{Oda0} in the canonical formalism. 

We would like to consider a bound state problem constructed out of the following composite scalar operator:
%**   Comp-op   %%%%%%%%%%%%%%%%%%%%%%%%%%%%%%%%%%%%%%%%%%%%%%%%%%%%%%%%%
\begin{eqnarray}
{\cal{O}}_{\varphi^2} (x) \equiv \varphi (x)^2. 
\label{Comp-op}  
\end{eqnarray}
%%%%%%%%%%%%%%%%%%%%%%%%%%%%%%%%%%%%%%%%%%%%%%%%%%%%%%%%%%%%%%%%%%% 
This is the most plausible operator if a bound state is made from
ghost fields in case of $\epsilon = - 1$. Since the other composite operators are dealt with in the same way, we will focus
our attention on whether bound states are made from the operator ${\cal{O}}_{\varphi^2} (x)$ in Eq. (\ref{Comp-op})
or not within the framework of the canonical operator formalism.

According to the Gell-Mann-Low formula \cite{Gell-Low}, in the canonical operator formalism the correlation function 
of the composite operator ${\cal{O}}_{\varphi^2} (x)$ is defined as\footnote{See Refs. \cite{Nishijima, Zimmermann1, Zimmermann2} 
for composite field operators in more detail.}
%**   Corr-funct   %%%%%%%%%%%%%%%%%%%%%%%%%%%%%%%%%%%%%%%%%%%%%%%%%%%%%%%%%
\begin{eqnarray}
&{}& \langle {\cal{O}}_{\varphi^2} (x) {\cal{O}}_{\varphi^2} (y) \rangle 
\equiv \langle {\bf{0}} | T [{\cal{O}}_{\varphi^2} (x) {\cal{O}}_{\varphi^2} (y) ] | {\bf{0}} \rangle 
\nonumber\\
&\equiv& \sum_{n=0}^\infty \frac{i^n}{n!} \lim_{\varepsilon \rightarrow +0} 
\int d^4 x_1 \cdots d^4 x_n \, {\rm{exp}} (- \varepsilon \sum_{i=1}^n | x_i^0 |)
\frac{1}{\langle 0 | S | 0 \rangle}
\nonumber\\
&\times& \langle 0 | T [ {\cal{L}}_{\rm{int}}(x_1) \cdots {\cal{L}}_{\rm{int}}(x_n) 
\varphi (x)^2 \varphi (y)^2 ] | 0 \rangle, 
\label{Corr-funct}  
\end{eqnarray}
%%%%%%%%%%%%%%%%%%%%%%%%%%%%%%%%%%%%%%%%%%%%%%%%%%%%%%%%%%%%%%%%%%% 
where a generic field operator ${\bf{\Phi}}(x)$ in the interaction picture is related to $\Phi(x)$ in the
Schrodinger picture by ${\bf{\Phi}}(x) = e^{iHx^0} \Phi(0, \Vec{x}) e^{-iHx^0}$ with $H$ being the 
Hamiltonian. Moreover, $| {\bf{0}} \rangle$ is the true vacuum which is different from the vacuum of the free
theory, $| 0 \rangle$ and $S$ is the S-matrix.\footnote{$| {\bf{0}} \rangle$ is related to $| 0 \rangle$ by 
$| {\bf{0}} \rangle = e^{i \theta} U(0, -\infty)| 0 \rangle$ where $\theta$ is a phase factor and $U$ is the standard U-matrix, which
is symbolically denoted as $U(t, t_0) \equiv T \exp \left(i \int_{t_0}^t d t^\prime \, {\cal{L}}^{\rm{int}}(t^\prime)
\right)$. Furthermore, $\langle 0 | S |  0 \rangle$ corresponds to the sum of vacuum polarization diagrams. The division by $\langle 0 | S |  0 \rangle$ 
means that one can neglect the Feynman diagrams involving the vacuum polarization.}

Following the formula (\ref{Corr-funct}), we can expand the correlation function in terms of the order of the coupling constant $\lambda$.
Using the Wick theorem, a lot of terms in general emerge. For instance, at the lowest order of $\lambda$, we have
%**   Order-0   %%%%%%%%%%%%%%%%%%%%%%%%%%%%%%%%%%%%%%%%%%%%%%%%%%%%%%%%%
\begin{eqnarray}
\langle {\cal{O}}_{\varphi^2} (x) {\cal{O}}_{\varphi^2} (y) \rangle \Big|_{ {\cal{O}}(\lambda^0) } 
= 2 D_\varphi (x-y)^2. 
\label{Order-0}  
\end{eqnarray}
%%%%%%%%%%%%%%%%%%%%%%%%%%%%%%%%%%%%%%%%%%%%%%%%%%%%%%%%%%%%%%%%%%% 
At the next leading order, we have
%**   Order-1   %%%%%%%%%%%%%%%%%%%%%%%%%%%%%%%%%%%%%%%%%%%%%%%%%%%%%%%%%
\begin{eqnarray}
&{}& \langle {\cal{O}}_{\varphi^2} (x) {\cal{O}}_{\varphi^2} (y) \rangle \Big|_{ {\cal{O}}(\lambda^1) } 
= - i \lambda \int d^4 z \, \Bigg[ \frac{1}{3} D_\varphi (x-y) D_\varphi (x-z) D_\varphi (y-z) D_\varphi (0) 
\nonumber\\
&+& D_\varphi (x-z)^2 D_\varphi (y-z)^2 \Bigg].
\label{Order-1}  
\end{eqnarray}
%%%%%%%%%%%%%%%%%%%%%%%%%%%%%%%%%%%%%%%%%%%%%%%%%%%%%%%%%%%%%%%%%%% 
However, the first term does not give rise to a bound state so we neglect this term, which amounts to confining ourselves to
the ladder approximation. Then, we can obtain 
%**   Order-1-2   %%%%%%%%%%%%%%%%%%%%%%%%%%%%%%%%%%%%%%%%%%%%%%%%%%%%%%%%%
\begin{eqnarray}
\langle {\cal{O}}_{\varphi^2} (x) {\cal{O}}_{\varphi^2} (y) \rangle \Big|_{ {\cal{O}}(\lambda^1) } 
\approx - i \lambda \int d^4 z \, D_\varphi (x-z)^2 D_\varphi (y-z)^2.
\label{Order-1-2}  
\end{eqnarray}
%%%%%%%%%%%%%%%%%%%%%%%%%%%%%%%%%%%%%%%%%%%%%%%%%%%%%%%%%%%%%%%%%%% 
In a similar manner, under the assumption of the ladder approximation, we can obtain the approximate expression of the correlation function up to
the order ${\cal{O}}(\lambda^3)$:  
%**   Appr-Cor-fun2   %%%%%%%%%%%%%%%%%%%%%%%%%%%%%%%%%%%%%%%%%%%%%%%%%%%%%%%%%
\begin{eqnarray}
&{}& \langle {\cal{O}}_{\varphi^2} (x) {\cal{O}}_{\varphi^2} (y) \rangle 
\nonumber\\
&\approx& 2 D_\varphi (x-y)^2  - i \lambda \int d^4 z \, D_\varphi (x-z)^2 D_\varphi (y-z)^2
\nonumber\\
&-& \frac{\lambda^2}{2} \int d^4 z d^4 w \, D_\varphi (x-z)^2 D_\varphi (y-w)^2 D_\varphi (z-w)^2
\nonumber\\
&+& {\cal{O}}(\lambda^3). 
\label{Appr-Cor-fun2}  
\end{eqnarray}
%%%%%%%%%%%%%%%%%%%%%%%%%%%%%%%%%%%%%%%%%%%%%%%%%%%%%%%%%%%%%%%%%%% 

In such a simplified context, one can sum over all quantum corrections and get its concise expression. Let us note that in order to understand
the issue of bound states, it is necessary to take quantum corrections at all orders into consideration. Before doing so, let us rewrite
each term into a simpler form in order. The first term on the RHS of Eq. (\ref{Appr-Cor-fun2}), which is at ${\cal{O}} (\lambda^0)$, 
can be cast to the form:
%**   D-D   %%%%%%%%%%%%%%%%%%%%%%%%%%%%%%%%%%%%%%%%%%%%%%%%%%%%%%%%%
\begin{eqnarray}
D_\varphi (x-y)^2 &=& \int \frac{d^4 q}{i (2 \pi)^4} e^{i q \cdot (x-y)} D_\varphi (q) 
\int \frac{d^4 q^\prime}{i (2 \pi)^4} e^{i q^\prime \cdot (x-y)} D_\varphi (q^\prime) 
\nonumber\\
&=& \int \frac{d^4 k}{i (2 \pi)^4} e^{i k \cdot (x-y)}  \int \frac{d^4 q}{i (2 \pi)^4} D_\varphi (q) D_\varphi (k-q) 
\nonumber\\
&\equiv& \int \frac{d^4 k}{i (2 \pi)^4} e^{i k \cdot (x-y)}  G(k), 
\label{D-D}  
\end{eqnarray}
%%%%%%%%%%%%%%%%%%%%%%%%%%%%%%%%%%%%%%%%%%%%%%%%%%%%%%%%%%%%%%%%%%% 
where $G(k)$ is defined by
%**   Def-G   %%%%%%%%%%%%%%%%%%%%%%%%%%%%%%%%%%%%%%%%%%%%%%%%%%%%%%%%%
\begin{eqnarray}
G(k) \equiv \int \frac{d^4 q}{i (2 \pi)^4} D_\varphi (q) D_\varphi (k-q).
\label{Def-G}  
\end{eqnarray}
%%%%%%%%%%%%%%%%%%%%%%%%%%%%%%%%%%%%%%%%%%%%%%%%%%%%%%%%%%%%%%%%%%% 
  
In a similar manner, we can rewrite the second and third terms on the RHS of Eq. (\ref{Appr-Cor-fun2}) whose result is simply given by
%** O(lambda)  %%%%%%%%%%%%%%%%%%%%%%%
\begin{eqnarray}
&{}& \int d^4 z \, D_\varphi (x-z)^2 D_\varphi (z-y)^2
= \int \frac{d^4 k}{i (2 \pi)^4} e^{i k \cdot (x - y)} \frac{1}{i} G(k)^2, 
\nonumber\\ \notag
&{}& \int d^4 z d^4 w \, D_\varphi (x-z)^2 D_\varphi (z-w)^2 D_\varphi (w-y)^2
= - \int \frac{d^4 k}{i (2 \pi)^4} e^{i k \cdot (x - y)} G(k)^3.
\\[0.5em]
\label{O(lambda)}  
\end{eqnarray}
%%%%%%%%%%%%%%%%%%%%%%%%%%%%%%%%%%%%%%%%%%%%%%%%%%%%%%%%%%%%%%%%%%% 
Substituting these expressions into (\ref{Appr-Cor-fun2}) leads to
%**   Appr-Cor-fun3   %%%%%%%%%%%%%%%%%%%%%%%%%%%%%%%%%%%%%%%%%%%%%%%%%%%%%%%%%
\begin{eqnarray}
&{}& \langle {\cal{O}}_{\varphi^2} (x) {\cal{O}}_{\varphi^2} (y) \rangle 
\equiv \int \frac{d^4 k}{i (2 \pi)^4} e^{i k \cdot (x-y)} C(k)
\nonumber\\
&=& \int \frac{d^4 k}{i (2 \pi)^4} e^{i k \cdot (x-y)} \, 2 G(k) \left[ 1 -  \frac{\lambda}{2} G(k) 
+ \frac{\lambda^2}{4} G(k)^2 +  {\cal{O}}(\lambda^3) \right]
\nonumber\\
&=& \int \frac{d^4 k}{i (2 \pi)^4} e^{i k \cdot (x-y)} \frac{2 G(k)}{1 +  \frac{\lambda}{2} G(k)},
\label{Appr-Cor-fun3}  
\end{eqnarray}
%%%%%%%%%%%%%%%%%%%%%%%%%%%%%%%%%%%%%%%%%%%%%%%%%%%%%%%%%%%%%%%%%%% 
from which we find that $C(p)$ takes the form:
%**   C(p)   %%%%%%%%%%%%%%%%%%%%%%%%%%%%%%%%%%%%%%%%%%%%%%%%%%%%%%%%%
\begin{eqnarray}
C(p) = \frac{2 G(p)}{1 +  \frac{\lambda}{2} G(p)}. 
\label{C(p)}  
\end{eqnarray}
%%%%%%%%%%%%%%%%%%%%%%%%%%%%%%%%%%%%%%%%%%%%%%%%%%%%%%%%%%%%%%%%%%% 
This equation implies that if there were a pole at some on-shell value, $p^2 = - {\cal{M}}^2$, we would have a bound state 
with the mass ${\cal{M}}$. Thus, by solving this equation in the next section, we will investigate the problem of whether there is such a bound state or not.

\section{Bound states}

As explained in the previous section, when the pole equation
%**   Pole-eq  %%%%%%%%%%%%%%%%%%%%%%%%%%%%%%%%%%%%%%%%%%%%%%%%%%%%%%%%%
\begin{eqnarray}
1 +  \frac{\lambda}{2} G(p) = 0,
\label{Pole-eq}  
\end{eqnarray}
%%%%%%%%%%%%%%%%%%%%%%%%%%%%%%%%%%%%%%%%%%%%%%%%%%%%%%%%%%%%%%%%%%% 
has a solution at the on-shell value, $p^2 = - {\cal{M}}^2$, one can judge that there is a bound state with the mass ${\cal{M}}$. Furthermore,
when the bound state has a positive norm, it is a physical bound state whereas when the bound state has a negative norm, 
it is a unphysical ghost bound state. 

For this aim, let us explicitly calculate $G(p)$ which has the logarithmic divergence.  From the definitions (\ref{D-mom}) and (\ref{Def-G}),
$G(p)$ is of form:
%**   G(p)  %%%%%%%%%%%%%%%%%%%%%%%%%%%%%%%%%%%%%%%%%%%%%%%%%%%%%%%%%
\begin{eqnarray}
G(p) &=& \int \frac{d^4 q}{i (2 \pi)^4} \, D_\varphi (q) D_\varphi (p-q)
\nonumber\\
&=& \int \frac{d^4 q}{i (2 \pi)^4} \, \frac{1}{(q^2 + m_\varphi^2 - i \varepsilon) [ (p-q)^2 + m_\varphi^2 - i \varepsilon ]}.
\label{G(p)}  
\end{eqnarray}
%%%%%%%%%%%%%%%%%%%%%%%%%%%%%%%%%%%%%%%%%%%%%%%%%%%%%%%%%%%%%%%%%%% 
In order to calculate this integral, we shall use the Pauli-Villars regularization. First, let us make use of the Feynman parameter formula:
%**   Feyn-para  %%%%%%%%%%%%%%%%%%%%%%%%%%%%%%%%%%%%%%%%%%%%%%%%%%%%%%%%%
\begin{eqnarray}
\frac{1}{A B} = \int_0^1 d x \, \frac{1}{[ A + (B - A) x ]^2}.
\label{Feyn-para}  
\end{eqnarray}
%%%%%%%%%%%%%%%%%%%%%%%%%%%%%%%%%%%%%%%%%%%%%%%%%%%%%%%%%%%%%%%%%%% 
Using this formula, $G(p)$ reads\footnote{Here we have shifted the variable of integration in momentum space, 
$q \rightarrow q + x p$. Strictly speaking, this procedure is only valid in convergent integrals. To justify the shift
of variables, it is necessary to introduce some regularization schemes such as the Pauli-Villars and dimensional ones.} 
%**   G(p)-2  %%%%%%%%%%%%%%%%%%%%%%%%%%%%%%%%%%%%%%%%%%%%%%%%%%%%%%%%%
\begin{eqnarray}
G(p) = \int \frac{d^4 q}{i (2 \pi)^4} \int_0^1 d x \, \frac{1}{(q^2 + \Delta )^2},
\label{G(p)-2}  
\end{eqnarray}
%%%%%%%%%%%%%%%%%%%%%%%%%%%%%%%%%%%%%%%%%%%%%%%%%%%%%%%%%%%%%%%%%%% 
where $\Delta$ is defined as
%**   Delta  %%%%%%%%%%%%%%%%%%%%%%%%%%%%%%%%%%%%%%%%%%%%%%%%%%%%%%%%%
\begin{eqnarray}
\Delta = p^2 x (1 - x) + m_\varphi^2 - i \epsilon.
\label{Delta}  
\end{eqnarray}
%%%%%%%%%%%%%%%%%%%%%%%%%%%%%%%%%%%%%%%%%%%%%%%%%%%%%%%%%%%%%%%%%%% 

As the next step, let us perform the Wick rotation. As long as $- p^2 < 4 m_\varphi^2$,  the quantity in $\Delta$, 
$p^2 x (1 - x) + m_\varphi^2$ is positive for all $x$ between $0$ and $1$, so the poles in the integrand of
Eq. (\ref{G(p)-2}) are located at 
%**   Wick-pole  %%%%%%%%%%%%%%%%%%%%%%%%%%%%%%%%%%%%%%%%%%%%%%%%%%%%%%%%%
\begin{eqnarray}
q^0 = \pm \sqrt{\Vec{q}\,^2 + p^2 x (1 - x) + m_\varphi^2 - i \epsilon}.
\label{Wick-pole}  
\end{eqnarray}
%%%%%%%%%%%%%%%%%%%%%%%%%%%%%%%%%%%%%%%%%%%%%%%%%%%%%%%%%%%%%%%%%%% 
Then, we can rotate the contour of integrations of $q^0$ counterclockwise without meeting the poles as usual,
so by putting $q^0 = i q^4$ we can integrate $q^0$ on the imaginary axis from $- i \infty$ to $+ i \infty$ instead of
integrating it on the real axis from $- \infty$ to $+ \infty$.  However, here a subtlety occurs.  
We are now interested in the case of $- p^2 > 4 m_\varphi^2$, so the quantity $p^2 x (1 - x) + m_\varphi^2$ 
is not always positive for all $x$ between $0$ and $1$. In this article, we assume that although this quantity
is not positive, the quantity $\Vec{q}\,^2 + p^2 x (1 - x) + m_\varphi^2$ is positive for all $x$ between $0$ and $1$, 
so that we can also perform the conventional Wick rotation. 

With $q^0 = i q^4$, we have $q^2 = - (q^0)^2 + \Vec{q}\,^2 = (q^4)^2 + \Vec{q}\,^2 \equiv q_E^2$ and
$\int d^4 q = i \int d^4 q_E$. Thus, adding the Pauli-Villars regulator, $G(p)$ can be rewritten in the Euclidean 
space-time as
%**   G(p)-E  %%%%%%%%%%%%%%%%%%%%%%%%%%%%%%%%%%%%%%%%%%%%%%%%%%%%%%%%%
\begin{eqnarray}
G(p) = \int_0^1 d x \int \frac{d^4 q_E}{(2 \pi)^4} \, \left[ \frac{1}{(q_E^2 + \Delta )^2} 
- \frac{1}{(q_E^2 + \Lambda^2)^2} \right],
\label{G(p)-E}  
\end{eqnarray}
%%%%%%%%%%%%%%%%%%%%%%%%%%%%%%%%%%%%%%%%%%%%%%%%%%%%%%%%%%%%%%%%%%% 
where $\Lambda$ is the cutoff. The integral over $q_E$ is easily computed to be\footnote{When we use the dimensional regularization 
instead of the Pauli-Villars regularization in $n$-dimensional space-time, $\log \Lambda^2$ is replaced with $\frac{1}{\bar \epsilon}
\equiv \frac{2}{4-n} + \log ( 4 \pi e^{-\gamma} )$ where $\gamma$ is the Euler number.}   
%**   G(p)-E2  %%%%%%%%%%%%%%%%%%%%%%%%%%%%%%%%%%%%%%%%%%%%%%%%%%%%%%%%%
\begin{eqnarray}
G(p) = \frac{1}{16 \pi^2} \int_0^1 d x \, \log \frac{\Lambda^2}{p^2 x (1 - x) + m_\varphi^2}. 
\label{G(p)-E2}  
\end{eqnarray}
%%%%%%%%%%%%%%%%%%%%%%%%%%%%%%%%%%%%%%%%%%%%%%%%%%%%%%%%%%%%%%%%%%% 
Though $G(p)$ is logarithmically divergent and dependent on the cutoff $\Lambda$, we can make it be a finite
value through renormalization of the field and coupling constant, and the choice of a suitable renormalization 
condition\footnote{See Appendix A for this derivation.}:
%**   G(p)-finite  %%%%%%%%%%%%%%%%%%%%%%%%%%%%%%%%%%%%%%%%%%%%%%%%%%%%%%%%%
\begin{eqnarray}
&{}& \frac{\lambda}{2} G({\cal{M}}^2) = - \frac{\lambda_R}{32 \pi^2} \int_0^1 d x \, \log 
\left[ 1 - \frac{{\cal{M}}^2 x (1 - x)}{m_\varphi^2} \right]
\nonumber\\
&=& - \frac{\lambda_R}{16 \pi^2} \Bigg\{ \sqrt{1 - \Big( \frac{2 m_\varphi}{{\cal{M}}} \Big)^2}
\log \Bigg[ \frac{{\cal{M}}}{2 m_\varphi} + \sqrt{\Big( \frac{{\cal{M}}}{2 m_\varphi} \Big)^2 - 1} \, \Bigg]
- 1 \Bigg\}, 
\label{G(p)-finite}  
\end{eqnarray}
%%%%%%%%%%%%%%%%%%%%%%%%%%%%%%%%%%%%%%%%%%%%%%%%%%%%%%%%%%%%%%%%%%% 
where $\lambda_R$ denotes a renormalized coupling constant and we have used the on-shell
condition, $p^2 = - {\cal{M}}^2$.

Hence the pole equation (\ref{Pole-eq}), which determines the (non)existence of bound states,
is given by
%**   Pole-eq2  %%%%%%%%%%%%%%%%%%%%%%%%%%%%%%%%%%%%%%%%%%%%%%%%%%%%%%%%%
\begin{eqnarray}
\sqrt{1 - \frac{1}{z^2}} \cosh^{-1} z = 1 + \frac{16 \pi^2}{\lambda_R}, 
\label{Pole-eq2}  
\end{eqnarray}
%%%%%%%%%%%%%%%%%%%%%%%%%%%%%%%%%%%%%%%%%%%%%%%%%%%%%%%%%%%%%%%%%%%
where we have put $z = \frac{{\cal{M}}}{2 m_\varphi} > 1$ and $\cosh^{-1} z \equiv \log ( z + \sqrt{z^2 -1} )$.
Let us set $f(z) \equiv \sqrt{1 - \frac{1}{z^2}} \cosh^{-1} z$. This function $f(z)$ is a monotonically increasing function in the region
of $1 \leq z \leq \infty$ where $f(1) = 0$. As a consistency check, let us consider the limit of $\lambda_R \rightarrow \infty$, which
means an infinitely strong coupling constant. Then, the RHS of Eq. (\ref{Pole-eq2}) approaches the value of $1$, so the solution exists
around $z \approx 2$, which means the existence of the bound state as physically expected. On the other hand, in the infinitely weak coupling 
constant limit $\lambda_R \rightarrow 0$, we do not expect the existence of a bound state. Actually, in this limit the RHS of Eq. (\ref{Pole-eq2}) 
approaches $\infty$, for which there is a solution when $z = \infty$, but $z$ must be finite so there is no solution, i.e., no bound state
as expected. 

As a physically plausible coupling constant, for instance, let us take $\lambda_R = 16 \pi^2$, for which the RHS of Eq. (\ref{Pole-eq2}) 
takes the value of $2$. In this case, there is a solution around $z \approx 4$. As this example and the other reasonable examples show,
the pole equation (\ref{Pole-eq2}) has a solution and we can therefore conclude that the composite operator ${\cal{O}}_{\varphi^2} (x) 
= \varphi (x)^2$ in fact makes a bound state.

To close this section, a few important remarks are in order. Firstly, in the above derivation, we have assumed $- p^2 > 4 m_\varphi^2$.
Let us now consider the opposite case, i.e., $- p^2 < 4 m_\varphi^2$\footnote{The specific case, $- p^2 = 4 m_\varphi^2$ would 
correspond to not a bound state but a resonant state as in the Nambu-Jona-Lasinio model \cite{Nambu}.}, for which we can perform 
the Wick rotation without any assumption. By following the similar analysis to the previous case, one can derive the equation:
%**   G(p)-finite-new  %%%%%%%%%%%%%%%%%%%%%%%%%%%%%%%%%%%%%%%%%%%%%%%%%%%%%%%%%
\begin{eqnarray}
\frac{\lambda}{2} G({\cal{M}}^2) = - \frac{\lambda_R}{16 \pi^2} \left[ \sqrt{\Big(\frac{2 m_\varphi}{{\cal{M}}}\Big)^2 - 1} \,
\tan^{-1} \left( \frac{1}{\sqrt{\Big(\frac{2 m_\varphi}{{\cal{M}}}\Big)^2 - 1}} \right) - 1 \right].
\label{G(p)-finite-new}  
\end{eqnarray}
%%%%%%%%%%%%%%%%%%%%%%%%%%%%%%%%%%%%%%%%%%%%%%%%%%%%%%%%%%%%%%%%%%% 
Then the pole equation reads
 %**   Pole-eq2-new  %%%%%%%%%%%%%%%%%%%%%%%%%%%%%%%%%%%%%%%%%%%%%%%%%%%%%%%%%
\begin{eqnarray}
\sqrt{\tilde z^2 - 1} \, \tan^{-1} \left( \frac{1}{\sqrt{\tilde z^2 - 1}} \right) = 1 + \frac{16 \pi^2}{\lambda_R}, 
\label{Pole-eq2-new}  
\end{eqnarray}
%%%%%%%%%%%%%%%%%%%%%%%%%%%%%%%%%%%%%%%%%%%%%%%%%%%%%%%%%%%%%%%%%%%
where we have defined $\tilde z = \frac{2 m_\varphi}{{\cal{M}}} > 1$. As before, it is easy to show that this equation is self-consistent and
has a solution exhibiting the existence of a bound state.

Secondly, we should point out that the bound state constructed from the composite 
operator $\varphi (x)^2$ is a normal bound state with positive norm. This fact can be easily seen from the following observation: When there is
a bound state in the channel of ${\cal{O}}_{\varphi^2} (x) = \varphi (x)^2$, the state has the same norm as the state, $(a_\varphi^\dagger)^2 
| 0 \rangle$. The norm of this state is positive since we can show that
%**   Comp-norm  %%%%%%%%%%%%%%%%%%%%%%%%%%%%%%%%%%%%%%%%%%%%%%%%%%%%%%%%%
\begin{eqnarray}
|| (a_\varphi^\dagger)^2 | 0 \rangle ||^2 = \langle 0 | a_\varphi a_\varphi a_\varphi^\dagger a_\varphi^\dagger | 0 \rangle
= 2 \epsilon^2 \langle 0 | 0 \rangle = 2, 
\label{Comp-norm}  
\end{eqnarray}
%%%%%%%%%%%%%%%%%%%%%%%%%%%%%%%%%%%%%%%%%%%%%%%%%%%%%%%%%%%%%%%%%%%
where the commutation relation (\ref{CRs}) was used.

Thirdly, as can be seen in the argument done thus far, various final results are completely independent of the sign factor $\epsilon$
although it appears at the intermediate stage of the calculation, so the existence of the bound state does not depend on 
whether the particle $\varphi$ is a normal particle or ghost. This fact can be understood from the fact that interactions are in general blind to 
the norm of particles.

Finally, it is interesting to consider whether the light normal particle $\phi$ also makes a bound state
or not. For this purpose, we will choose ${\cal{O}}_{\phi^2} (x) = \phi (x)^2$ as the composite operator. Using the ladder approximation again, 
it turns out that the pole equation remains the same as Eq. (\ref{Pole-eq2}) except for the replacement of $z = \frac{{\cal{M}}}{2 m_\varphi}$ 
by $w \equiv \frac{{\cal{M}}}{2 m_\phi}$.     
Here let us suppose the situation where two massless gravitons with $m_\phi \rightarrow 0$ make a bound state with a very 
tiny but finite mass ${\cal{M}}$. In this case, we have $w \gg 1$ so there is no solution to the pole equation except for the unrealistic
coupling constant $\lambda_R \approx 0$. In this sense, the light and normal particle $\phi$ does not make a bound state. In relation to quadratic gravity,
it is therefore worthwhile to stress that the massive ghost has a tendency of making a bound state whereas the massless graviton does not so
in the theory with the quartic interaction term.

\section{Relationship with quadratic gravity}

In this section, we wish to simply point out the relationship between the fourth-derivative scalar theory and quadratic gravity.
The detail of the calculation will be postponed in the future work.  

Let us begin by an action of quadratic gravity using the conformal tensor\footnote{We follow the notation and conventions of 
Misner-Thorne-Wheeler (MTW) textbook \cite{MTW}. }:
%**   QG-action   %%%%%%%%%%%%%%%%%%%%%%%%%%%%%%%%%%%%%%%%%%%%%%%%%%%%%%%%%
\begin{eqnarray}
S = \int d^4 x \, \sqrt{-g} \left( \frac{1}{2 \kappa^2} R - \alpha C_{\mu\nu\rho\sigma}^2 + \beta R^2 \right),
\label{QG-action}  
\end{eqnarray}
%%%%%%%%%%%%%%%%%%%%%%%%%%%%%%%%%%%%%%%%%%%%%%%%%%%%%%%%%%%%%%%%%%% 
where $\kappa^2 \equiv 8 \pi G$ with $G$ being the Newton constant, $R$ is the scalar curvature and $C_{\mu\nu\rho\sigma}$ is 
the conformal tensor which defined by
%**   C-tensor   %%%%%%%%%%%%%%%%%%%%%%%%%%%%%%%%%%%%%%%%%%%%%%%%%%%%%%%%%
\begin{eqnarray}
C_{\mu\nu\rho\sigma} &=& R_{\mu\nu\rho\sigma} - \frac{1}{2} ( g_{\mu\rho} R_{\nu\sigma}
- g_{\mu\sigma} R_{\nu\rho} - g_{\nu\rho} R_{\mu\sigma} + g_{\nu\sigma} R_{\mu\rho} )
\nonumber\\
&+& \frac{1}{6} ( g_{\mu\rho} g_{\nu\sigma} - g_{\mu\sigma} g_{\nu\rho} ) R.
\label{C-tensor}  
\end{eqnarray}
%%%%%%%%%%%%%%%%%%%%%%%%%%%%%%%%%%%%%%%%%%%%%%%%%%%%%%%%%%%%%%%%%%% 
Moreover, the coupling constants $\alpha$ and $\beta$ are positive and dimensionless. Also recall that these coupling constants
are asymptotically free \cite{Mario, Fradkin, Avramidi}.

In the weak field approximation we can expand
%**   eta-exp   %%%%%%%%%%%%%%%%%%%%%%%%%%%%%%%%%%%%%%%%%%%%%%%%%%%%%%%%%
\begin{eqnarray}
g_{\mu\nu} = \eta_{\mu\nu} + h_{\mu\nu},
\label{eta-exp}  
\end{eqnarray}
%%%%%%%%%%%%%%%%%%%%%%%%%%%%%%%%%%%%%%%%%%%%%%%%%%%%%%%%%%%%%%%%%%% 
where $| h_{\mu\nu} | \ll 1$.  With this expansion, quadratic terms with respect to $h_{\mu\nu}$ in the action (\ref{QG-action}) take the form:  
%**   QG-action2   %%%%%%%%%%%%%%%%%%%%%%%%%%%%%%%%%%%%%%%%%%%%%%%%%%%%%%%%%
\begin{eqnarray}
S = \frac{1}{4 \kappa^2} \int d^4 x \, \left[ - \frac{1}{2 m_1^2} h^{\mu\nu} (\Box -  m_1^2) P_{\mu\nu, \rho\sigma}^{(2)} \Box h^{\rho\sigma}
 + \frac{1}{m_2^2} h^{\mu\nu} (\Box -  m_2^2) P_{\mu\nu, \rho\sigma}^{(0, s)} \Box h^{\rho\sigma} \right].
 \notag
 \\[0.5em]
\label{QG-action2}  
\end{eqnarray}
%%%%%%%%%%%%%%%%%%%%%%%%%%%%%%%%%%%%%%%%%%%%%%%%%%%%%%%%%%%%%%%%%%% 
Here $P_{\mu\nu, \rho\sigma}^{(2)}$ and $P_{\mu\nu, \rho\sigma}^{(0, s)}$ are projectors to the spin-$2$ and spin-$0$ states, respectively
\cite{Percacci, Nakasone}, and $m_1^2 \equiv \frac{1}{4 \kappa^2 \alpha}, m_2^2 \equiv \frac{1}{12 \kappa^2 \beta}$.  
  
First, let us pay our attention to the second term on the RHS of Eq. (\ref{QG-action2}) including the projector $P_{\mu\nu, \rho\sigma}^{(0, s)}$.  
At first sight, it appears that this term gives rise to a massless ghost with negative norm, but this ghost turns out be gauged away by
the general coordinate symmetry \cite{Oda-f}. As a result, there remains only a physical massive particle with positive norm, which is sometimes 
called ``scalaron'' in the context of cosmology.

Next, we move on to the first term on the RHS of Eq. (\ref{QG-action2}) including the projector $P_{\mu\nu, \rho\sigma}^{(2)}$, which
describes the spin-$2$ tensor sector. Without gauge fixing, the propagator in this sector can be obtained and given in momentum space:
%**   QG-prop   %%%%%%%%%%%%%%%%%%%%%%%%%%%%%%%%%%%%%%%%%%%%%%%%%%%%%%%%%
\begin{eqnarray}
\frac{4 \kappa^2 m_1^2}{p^2 ( p^2 + m_1^2)} P_{\mu\nu, \rho\sigma}^{(2)} 
= 4 \kappa^2 \left( \frac{1}{p^2} - \frac{1}{p^2 + m_1^2} \right) P_{\mu\nu, \rho\sigma}^{(2)}.
 \notag
 \\[0.5em]
\label{QG-prop}  
\end{eqnarray}
%%%%%%%%%%%%%%%%%%%%%%%%%%%%%%%%%%%%%%%%%%%%%%%%%%%%%%%%%%%%%%%%%%% 
This expression shows that quadratic gravity possesses two spin-$2$ degrees of freedom, one of which is a massless spin-$2$ particle
with positive norm, which can be identified with the graviton, and the other is a massive spin-$2$ particle with negative norm,
which is called the massive ghost and provides us with a pathology. In this way, we have demonstrated that quadratic gravity 
shares common features with the scalar theory with fourth-order derivative term at least in the kinetic terms.

If we explore the analogy between the two theories further, it is natural to ask ourselves whether there is a $\Phi^4$ interaction
term in quadratic gravity as in the scalar theory. Since we now assume that the cosmological constant is vanishing, 
there is not such a term in quadratic gravity. However, for instance, from $\sqrt{-g} R^2$ in the action (\ref{QG-action})     
we have a term proportional to $\beta (h^{\mu\nu} P_{\mu\nu, \rho\sigma}^{(2)} \Box h^{\rho\sigma})^2$. This term gives us
the $\Phi^4$ interaction with the momentum-dependent coupling constant in momentum space.

\section{Conclusions}

In this paper, based on the canonical operator formalism, we have demonstrated that there is a bound state in the scalar theory with
fourth-order derivative term. We should stress that we have only made use of not path integral but the operator formalism, which is 
the basic tool of QFTs with sound foundation. Moreover, we have shown that the massive ghost, which violates unitarity, has a tendency
for forming the bound state while the almost massless normal particle, would correspond to the graviton, does not so. 

One of our motivations behind the present study is to apply the formalism at hand to the problem of unitarity violation arising from
the massive ghost in quadratic gravity. As shown in the last section, the scalar theory with fourth-order derivative term has a similar
structure to quadratic gravity, so it is expected that the results obtained in this paper are valid even in quadratic gravity to some extent .

It is a pity that although a pair of the massive ghosts forms a bound state with positive norm, we cannot solve the problem of the
unitarity violation due to the massive ghost in quadratic gravity since the bound state is dissolved into elementary ghost fields 
in the weak coupling regime and consequently the problem of the unitarity violation reappears. In order to solve the problem from 
the viewpoint of bound states, we need a permanent confinement of the massive ghost like the confinement of quarks and gluons in QCD
or the confinement of the Faddeev-Popov (FP) ghosts in any gauge theories.
 
Under such a situation, it is valuable to pursue a new solution to the massive ghost problem from the viewpoint of bound states 
within the framework of the canonical formalism. For instance, in the analysis of physical modes of 
quadratic gravity \cite{Oda-Can}\footnote{The BRST formalism of various gravitational theories has been already constructed 
in the de Donder (harmonic) gauge \cite{Oda-f, Oda-Q, Oda-W, Oda-Saake, Oda-Corfu, Oda-Ohta, Oda-Conf}.}, 
we usually assume that there is no bound state. If this assumption is relaxed, we are free to take account of various kinds of bound states 
which might be constructed from the massive ghost and the other elementary fields such as the FP ghosts.

In this respect, in QCD for which the confinement of quarks and gluons occurs, it is encouraging to notice that there
appear a lot of bound states in a natural way. For instance, the BRST transformation of quark field $\psi$ takes the form:
%**   BRST-quark   %%%%%%%%%%%%%%%%%%%%%%%%%%%%%%%%%%%%%%%%%%%%%%%%%%%%%%%%%
\begin{eqnarray}
\{ Q_B, \psi (x) \} = g c(x) \psi (x),
\label{BRST-quark}  
\end{eqnarray}
%%%%%%%%%%%%%%%%%%%%%%%%%%%%%%%%%%%%%%%%%%%%%%%%%%%%%%%%%%%%%%%%%%% 
where $Q_B$ is the BRST charge corresponding to the colour gauge transformation, $g$ the coupling constant, 
and $c$ is the FP ghost field  \cite{Kugo-Ojima}.
Since the quark $\psi$ has a color, it is not BRST invariant but belongs to a BRST doublet. This fact implies that the composite operator
$c(x) \psi (x)$ must form a bound state ${\cal {C}} (x)$, and as a result we have a BRST doublet, $( \psi (x), {\cal {C}} (x) )$. In a similar
manner, the BRST transformation of the gauge field requires the composite field $A (x) \times c(x)$ to form a bound state.  This situation 
encourages us to examine the issue of bound states in quadratic gravity. If a certain bound state is constructed from the massive ghost 
and the FP ghosts etc., and its BRST-conjugate corresponding to this bound state is also formed, they together constitute a BRST doublet, 
thereby making it possible to solve the massive ghost problem \cite{Kawasaki} since this mechanism realizes the scenario of 
the permanent confinement of the massive ghost via the BRST quartet mechanism \cite{Kugo-Ojima}. We wish to return to this problem in the near future.

%%%%%%%%%%%%%%%%%%%%%%%% Acknowledgements %%%%%%%%%%%%%%%%%%%%%%%%%%%%%
%%%%%%%%%%%%%%%%%%%%%%%%%%%%%%%%%%%%%%%%%%%%%%%%%%%%%%%%%%%%%%%%%%
\begin{flushleft}
{\bf Acknowledgements}
\end{flushleft}

We would like to thank T. Kugo for valuable discussions.

%%%%%%%%%%%%%%%%%%%%%%% Appendix %%%%%%%%%%%%%%%%%%%%%%%%%%%%%%%
%%%%%%%%%%%%%%%%%%%%%%%%%%%%%%%%%%%%%%%%%%%%%%%%%%%%%%%%%%%%%%%%%%
\appendix
\addcontentsline{toc}{section}{Appendix~\ref{app:scripts}: Training Scripts}
\section*{Appendix}
\label{app:scripts}
\renewcommand{\theequation}{A.\arabic{equation}}
\setcounter{equation}{0}

\section{Derivation of Eq. (\ref{G(p)-finite})}
\def\T{\text{T}}

In this appendix, we demonstrate a renormalization of the composite operator and present a derivation of 
Eq. (\ref{G(p)-finite}) \cite{Nishijima, Zimmermann1, Zimmermann2, Weinberg}.  

In order to renormalize $G(p)$ of Eq. (\ref{G(p)-E2}), let us consider a renormalization of the composite operator, ${\cal{O}}_{\varphi^2} (x)$.
To do that, we shall compute a matrix element, $\langle \beta | {\cal{O}}_{\varphi^2} (p) | \alpha \rangle$ of the composite operator in
momentum space, for which we define
%**   Com-moment   %%%%%%%%%%%%%%%%%%%%%%%%%%%%%%%%%%%%%%%%%%%%%%%%%%%%%%%%%
\begin{eqnarray}
{\cal{O}}_{\varphi^2} (p) = i \int d^4 x \, e^{- i p x} {\cal{O}}_{\varphi^2} (x) = i \int d^4 x \, e^{- i p x} \varphi (x)^2.  
\label{Com-moment}  
\end{eqnarray}
%%%%%%%%%%%%%%%%%%%%%%%%%%%%%%%%%%%%%%%%%%%%%%%%%%%%%%%%%%%%%%%%%%% 
In the language of Feynman diagrams, this corresponds to inserting a vertex where two internal $\varphi$-lines come together, and then
through which a total momentum $p$ enters the diagram where the transition of states, $\alpha \rightarrow \beta$ happens. 

The logarithmic divergence arises from a class of diagrams where this new vertex is part of a subdiagram which is connected to the rest
of the diagrams by two $\varphi$ internal lines. To the ${\cal{O}}(\lambda^2)$-order, the relevant subdiagram makes a contribution to
the matrix element of ${\cal{O}}_{\varphi^2} (p)$ a divergent factor   
%**   K(p)  %%%%%%%%%%%%%%%%%%%%%%%%%%%%%%%%%%%%%%%%%%%%%%%%%%%%%%%%%
\begin{eqnarray}
K(p) = 1 - \frac{\lambda}{2} \int \frac{d^4 q}{i (2 \pi)^4} \, \frac{1}{(q^2 + m_\varphi^2 - i \varepsilon) [ (p-q)^2 + m_\varphi^2 - i \varepsilon ]}.
\label{K(p)}  
\end{eqnarray}
%%%%%%%%%%%%%%%%%%%%%%%%%%%%%%%%%%%%%%%%%%%%%%%%%%%%%%%%%%%%%%%%%%% 
Note that this certainly coincides with the denominator of $C(p)$ in Eq. (\ref{C(p)}) because $\frac{1}{1 +  \frac{\lambda}{2} G(p)} = 1 - \frac{\lambda}{2}
G(p) + {\cal{O}}(\lambda^2)$ up to the ${\cal{O}}(\lambda^2)$-order.

Along the same line of argument as before, $K(p)$ is calculated to be
%**   K(p)-2  %%%%%%%%%%%%%%%%%%%%%%%%%%%%%%%%%%%%%%%%%%%%%%%%%%%%%%%%%
\begin{eqnarray}
K(p) = 1 - \frac{1}{32 \pi^2} \int_0^1 d x \, \log \frac{\Lambda^2}{p^2 x (1 - x) + m_\varphi^2} + {\cal{O}}(\lambda^2). 
\label{K(p)-2}  
\end{eqnarray}
%%%%%%%%%%%%%%%%%%%%%%%%%%%%%%%%%%%%%%%%%%%%%%%%%%%%%%%%%%%%%%%%%%% 
This logarithmic divergence can be eliminated by introducing a renormalized composite operator which is defined as
%**   Ren-op  %%%%%%%%%%%%%%%%%%%%%%%%%%%%%%%%%%%%%%%%%%%%%%%%%%%%%%%%%
\begin{eqnarray}
{\cal{O}}_{\varphi^2 \, R} (x) = Z \, \varphi (x)^2, 
\label{Ren-op}  
\end{eqnarray}
%%%%%%%%%%%%%%%%%%%%%%%%%%%%%%%%%%%%%%%%%%%%%%%%%%%%%%%%%%%%%%%%%%% 
where the renormalization factor $Z$ is chosen in such a way that $Z \, K(p)$ becomes finite at a certain renormalization point.
For instance, let us choose
%**   Ren-point  %%%%%%%%%%%%%%%%%%%%%%%%%%%%%%%%%%%%%%%%%%%%%%%%%%%%%%%%%
\begin{eqnarray}
Z \, K (0) = 1. 
\label{Ren-point}  
\end{eqnarray}
%%%%%%%%%%%%%%%%%%%%%%%%%%%%%%%%%%%%%%%%%%%%%%%%%%%%%%%%%%%%%%%%%%% 
Then, we can obtain 
%**   Z-factor  %%%%%%%%%%%%%%%%%%%%%%%%%%%%%%%%%%%%%%%%%%%%%%%%%%%%%%%%%
\begin{eqnarray}
Z = 1 + \frac{\lambda}{32 \pi^2} \int_0^1 d x \, \log \frac{\Lambda^2}{m_\varphi^2} + {\cal{O}}(\lambda^2). 
\label{Z-factor}  
\end{eqnarray}
%%%%%%%%%%%%%%%%%%%%%%%%%%%%%%%%%%%%%%%%%%%%%%%%%%%%%%%%%%%%%%%%%%% 
Thus, the matrix element of the renormalized composite operator ${\cal{O}}_{\varphi^2 \, R} (x)$ contains the factor:
%**   Renor-K(p)  %%%%%%%%%%%%%%%%%%%%%%%%%%%%%%%%%%%%%%%%%%%%%%%%%%%%%%%%%
\begin{eqnarray}
K_R (p) &\equiv& Z \, K(p)
\nonumber\\
&=& 1 + \frac{\lambda}{32 \pi^2} \int_0^1 d x \, \log \left[ 1 + \frac{p^2 x (1 - x)}{m_\varphi^2} \right] + {\cal{O}}(\lambda^2). 
\label{Renor-K(p)}  
\end{eqnarray}
%%%%%%%%%%%%%%%%%%%%%%%%%%%%%%%%%%%%%%%%%%%%%%%%%%%%%%%%%%%%%%%%%%% 
From the correspondence $\frac{1}{1 +  \frac{\lambda}{2} G(p)} = 1 - \frac{\lambda}{2} G(p) + {\cal{O}}(\lambda^2)$, we find the
expression on-shell $p^2 = - {\cal{M}}^2$:
%**   G(p)-finite2  %%%%%%%%%%%%%%%%%%%%%%%%%%%%%%%%%%%%%%%%%%%%%%%%%%%%%%%%%
\begin{eqnarray}
\frac{\lambda}{2} G({\cal{M}}^2) = - \frac{\lambda_R}{32 \pi^2} \int_0^1 d x \, \log 
\left[ 1 - \frac{{\cal{M}}^2 x (1 - x)}{m_\varphi^2} \right], 
\label{G(p)-finite2}  
\end{eqnarray}
%%%%%%%%%%%%%%%%%%%%%%%%%%%%%%%%%%%%%%%%%%%%%%%%%%%%%%%%%%%%%%%%%%% 
which is nothing but Eq. (\ref{G(p)-finite}). In order to eleminate the cutoff in higher-order calculations, we need to introduce 
the renormalized coupling constant $\lambda_R$ in addition to the field renormalization as well.

%%%%%%%%%%%%%%%%%%%%%%% reference %%%%%%%%%%%%%%%%%%%%%%%%%%%%%%%
%%%%%%%%%%%%%%%%%%%%%%%%%%%%%%%%%%%%%%%%%%%%%%%%%%%%%%%%%%%%%%%%%%

\end{document}